\documentclass[12pt]{article}

\usepackage{amsmath,bbm,latexsym}
\usepackage{epsfig}

\newcommand{\be}{\begin{equation}}
\newcommand{\ee}{\end{equation}}
\newcommand{\ba}{\begin{eqnarray}}
\newcommand{\ea}{\end{eqnarray}}
\newcommand{\no}{\nonumber\\}

\newcommand{\lesssim}{ \ \mbox{\raisebox{-3pt}{$\stackrel%
{\displaystyle <}{\sim}$}} \ }
\newcommand{\gtrsim}{\:\mbox{\raisebox{-3pt}{$\stackrel%
{\displaystyle >}{\sim}$}}\:}

\begin{document}

\title{
\normalsize \hfill CFTP/11-019 \\[1cm]
\LARGE Seesaw neutrino masses from an $A_4$ model \\
with two equal vacuum expectation values}

\author{P.~M.~Ferreira$^{(1,2)}$\thanks{E-mail: ferreira@cii.fc.ul.pt}
\ and L.~Lavoura$^{(3)}$\thanks{E-mail: balio@cftp.ist.utl.pt}
\\*[3mm]
\small $^{(1)}$ Instituto Superior de Engenharia de Lisboa,
\small 1959-007 Lisboa, Portugal
\\*[2mm]
\small $^{(2)}$
Centre of Theoretical and Computational Physics, University of Lisbon \\
\small 1649-003 Lisboa, Portugal
\\*[2mm]
\small $^{(3)}$ Technical University of Lisbon,
Centre of Theoretical Particle Physics \\
\small Instituto Superior T\'ecnico, 1049-001 Lisboa, Portugal
}

\date{8 March 2012}

\maketitle

\begin{abstract}
We present a model for the lepton sector,
with $A_4$ horizontal-symmetry group,
in which two of the Higgs doublets in an $A_4$ triplet of Higgs doublets
have equal vacuum expectation values.
The model makes well-defined predictions
for the effective light-neutrino Majorana mass matrix.
We show that those predictions are compatible with the experimental data.
\end{abstract}

\newpage

\section{Introduction}

Particle physics now boasts an impressive knowledge
of the three light-neutrino masses,
$m_{1,2,3}$,
and of lepton mixing.
The latter is parametrized by the mixing matrix
\be
U = \left( \begin{array}{ccc}
c_{12} c_{13} & s_{12} c_{13} & s_{13} e^{- i \delta} \\
- s_{12} c_{23} - c_{12} s_{23} s_{13} e^{i \delta} &
c_{12} c_{23} - s_{12} s_{23} s_{13} e^{i \delta} &
s_{23} c_{13} \\
s_{12} s_{23} - c_{12} c_{23} s_{13} e^{i \delta} &
- c_{12} s_{23} - s_{12} c_{23} s_{13} e^{i \delta} &
c_{23} c_{13}
\end{array} \right),
\ee
where $c_m = \cos{\theta_m}$ and $s_m = \sin{\theta_m}$
for $m \in \left\{ 12, 13, 23 \right\}$.
That knowledge is summarized in table~\ref{data},
\begin{table}[h]
\begin{center}
\renewcommand{\arraystretch}{1.2}
\begin{tabular}{|c|c|c|}
\hline
parameter & best-fit value & $3 \sigma$ interval
\\ \hline \hline
$m_2^2 - m_1^2$ (in $10^{-5}~\mathrm{eV}^2$)
& 7.59 & [7.09, 8.19]
\\ \hline
$\left| m_3^2 - m_1^2 \right|$ (in $10^{-3}~\mathrm{eV}^2$)
& $\begin{array}{c} 2.50 \\*[-2mm] 2.40 \end{array}$
& $\begin{array}{c} [ 2.14, 2.76 ] \\*[-1mm] [2.13, 2.67] \end{array}$
\\ \hline
$s_{12}^2$ & 0.312 & [0.27, 0.36]
\\ \hline
$s_{23}^2$ & 0.52 & [0.39, 0.64]
\\ \hline
$s_{13}^2$
& $\begin{array}{c} 0.013 \\*[-2mm] 0.016 \end{array}$
& $\begin{array}{c} [ 0.001, 0.035 ] \\*[-1mm] [ 0.001, 0.039 ] \end{array}$
\\ \hline
\end{tabular}
\end{center}
\caption{Experimental data for the neutrino masses and for lepton mixing.
In the cases of $\left| m_3^2 - m_1^2 \right|$ and of $s_{13}^2$,
the upper (lower) line corresponds to the case of a normal (inverted)
neutrino mass spectrum.
}
\label{data}
\end{table}
which is borrowed in abridged form
from reference~\cite{schwetz}.\footnote{An alternative
phenomenological fit to the data is given in reference~\cite{fogli}.}
Notice that we do not know the absolute mass scale of the neutrinos.
We also ignore whether the neutrino mass spectrum is `normal',
\textit{i.e.}\ with $m_3 > m_{1,2}$,
or `inverted',
\textit{i.e.}\ with $m_3 < m_{1,2}$.
Finally,
we lack much information on the phase $\delta$,
which remains essentially free.

Flavour physics would like to find a rationale for these experimental data
by imposing `horizontal' symmetries on the leptonic Lagrangian
and by assuming a pattern of
(spontaneous or soft)
breaking of those symmetries.
A review of some achievements in this field
can be found in reference~\cite{review}.
In particular,
a horizontal-symmetry group much used in this context
has been $A_4$~\cite{ma},
which is the smallest group
with a triplet irreducible representation.
Models using $A_4$~\cite{a4}
usually feature a horizontal-$A_4$ triplet
of scalar `Higgs' gauge-$SU(2)$ doublets;
in that triplet
either only one Higgs doublet has nonzero vacuum expectation value (VEV)
or all three Higgs doublets have equal VEVs.

This paper presents a new model
using horizontal-symmetry group $A_4$ in the lepton sector.
The model has the novel feature that in the $A_4$ triplet of Higgs doublets
\emph{two of the doublets have equal VEVs}
(different from the VEV of the third doublet).
Subsection~\ref{vacuum} explains how such a vacuum can come about.

The model that we suggest uses the type-I seesaw mechanism
and makes clear-cut predictions.
Let the symbol $M$ denote the effective Majorana
(hence symmetric)
light-neutrino mass matrix
in the basis where the charged-lepton mass matrix is diagonal.
Then those predictions are
\ba
M_{ee} M_{\mu\mu} M_{\tau\tau}
&=& M_{e\mu} M_{e\tau} M_{\mu\tau},
\label{constraint1}
\\
M_{\mu\mu} \left( M_{e\tau} \right)^2
&=& M_{\tau\tau} \left( M_{e\mu} \right)^2.
\label{constraint2}
\ea
These predictions are invariant under a rephasing of $M$,
\textit{i.e.}\ under the transformation
\be
M_{\alpha \alpha^\prime} \to
e^{i \left( \psi_\alpha + \psi_{\alpha^\prime} \right)}
M_{\alpha \alpha^\prime},
\label{reph}
\ee
for $\alpha, \alpha^\prime \in \left\{ e, \mu, \tau \right\}$.
They thus embody four real constraints on $M$,
just as when $M$ has either two vanishing matrix elements~\cite{marfatia}
or two vanishing minors~\cite{lavoura}.

In section~\ref{model} we present our model.
In section~\ref{predictions} we display its predictions
by means of scatter plots for the various observables.
A brief summary of our achievements is given in section~\ref{conclusions}.

\section{The model} \label{model}

\subsection{Fields and symmetries}

We envisage an extension of the Standard Model,
with gauge group $SU(2) \times U(1)$,
in which there are three right-handed neutrinos
$\nu_{1,2,3R}$ and four scalar $SU(2)$ doublets $\Phi_{1,2,3,4}$.
As usual,
there are three left-handed lepton $SU(2)$ doublets $D_{\alpha L}$
and three right-handed charged-lepton $SU(2)$ singlets $\alpha_R$.
The quark sector will not be dealt with in this paper,
but the model can in principle be extended to accomodate it.

The model has horizontal-symmetry group $A_4$.
The group $A_4$ is generated by two transformations,
$S$ and $T$.
Those transformations act in the following way:
\be
S: \quad
\nu_{2R} \to - \nu_{2R}, \
\nu_{3R} \to - \nu_{3R}, \
\Phi_2 \to - \Phi_2, \
\Phi_3 \to - \Phi_3;
\ee
\ba
T: & &
\nu_{1R} \to \nu_{2R} \to \nu_{3R} \to \nu_{1R}, \
\Phi_1 \to \Phi_2 \to \Phi_3 \to \Phi_1,
\no & &
D_{\mu L} \to \omega D_{\mu L}, \
D_{\tau L} \to \omega^2 D_{\tau L}, \
\mu_R \to \omega \mu_R, \
\tau_R \to \omega^2 \tau_R,
\ea
where $\omega = \exp \left( 2 i \pi / 3 \right)$.\footnote{Technically,
$\left( \nu_{1R}, \nu_{2R}, \nu_{3R} \right)$
and $\left( \Phi_1,  \Phi_2, \Phi_3 \right)$ are $\mathbf{3}$ of $A_4$,
$D_{\mu L}$ and $\mu_R$ are $\mathbf{1^\prime}$ of $A_4$,
and $D_{\tau L}$ and $\tau_R$ are $\mathbf{1^{\prime\prime}}$ of $A_4$.}
The Higgs doublet $\Phi_4$ and the lepton multiplets $D_{eL}$ and $e_R$
are $A_4$-invariant.\footnote{Further $A_4$-invariant Higgs doublets
could be used to give masses to the quarks.}

\subsection{Lagrangian}

Let $\Phi_j =
\left( \begin{array}{c} \phi_j^+ \\ \phi_j^0 \end{array} \right)$
($j \in \left\{ 1, 2, 3, 4 \right\}$)
and let $v_j$ denote the VEV of $\phi_j^0$.
The masses of the charged leptons originate in Yukawa couplings to $\Phi_4$:
\be
\mathcal{L}_{\ell \mathrm{Yukawa}}
= - \left( \sum_{\alpha = e, \mu, \tau}
\frac{m_\alpha}{v_4} \bar D_{\alpha L} \alpha_R \right)
\Phi_4 + \mathrm{H.c.};
\label{jbvut}
\ee
we may choose the phase of $\alpha_R$
in such a way that $m_\alpha$ is real and positive.
The charged-lepton mass matrix is automatically diagonal
because of the horizontal symmetry.

The Majorana mass terms of the right-handed neutrinos are
\be
\mathcal{L}_\mathrm{Maj} = m
\sum_{k=1}^3 \nu_{kR}^T C^{-1} \nu_{kR}
+ \mathrm{H.c.},
\ee
where $C$ is the charge-conjugation matrix in Dirac space.
Because of the horizontal symmetry,
$\mathcal{L}_\mathrm{Maj}$ is proportional to the unit matrix in flavour space.
Therefore, the effective light-neutrino Majorana mass matrix
following from the type-I seesaw mechanism is simply
\be
M = - \frac{1}{m} M_D M_D^T,
\label{M1}
\ee
where $M_D$ is the Dirac mass matrix connecting the left-handed
to the right-handed neutrinos.

The Yukawa couplings of the right-handed neutrinos are given by
\ba
\mathcal{L}_{\nu \mathrm{Yukawa}} &=&
a \bar D_{eL} \left( \nu_{1R} \tilde \Phi_1
+ \nu_{2R} \tilde \Phi_2 + \nu_{3R} \tilde \Phi_3 \right)
\no & &
+ b \bar D_{\mu L} \left( \nu_{1R} \tilde \Phi_1
+ \omega^2 \nu_{2R} \tilde \Phi_2 + \omega \nu_{3R} \tilde \Phi_3 \right)
\no & &
+ c \bar D_{\tau L} \left( \nu_{1R} \tilde \Phi_1
+ \omega \nu_{2R} \tilde \Phi_2 + \omega^2 \nu_{3R} \tilde \Phi_3 \right)
+ \mathrm{H.c.},
\label{jvuit}
\ea
%
where $\tilde \Phi_j \equiv \left( \begin{array}{cc}
0 & 1 \\ -1 & 0 \end{array} \right) \Phi_j^\ast
= \left( \begin{array}{c} {\phi_j^0}^\ast \\ - \phi_j^- \end{array} \right)$
and $a$,
$b$,
and $c$ are complex dimensionless coupling constants.
It follows from equation~(\ref{jvuit}) that
\be
M_D = \left( \begin{array}{ccc}
a v_1^\ast & a v_2^\ast & a v_3^\ast \\
b v_1^\ast & \omega^2 b v_2^\ast & \omega b v_3^\ast \\
c v_1^\ast & \omega c v_2^\ast & \omega^2 c v_3^\ast
\end{array} \right).
\label{uvhqw}
\ee

The scalar potential is
\ba
V &=&
\mu_1 \left( \Phi_1^\dagger \Phi_1 + \Phi_2^\dagger \Phi_2
+ \Phi_3^\dagger \Phi_3 \right)
+ \mu_2 \, \Phi_4^\dagger \Phi_4
\no & &
+ \lambda_1 \left[ \left( \Phi_1^\dagger \Phi_1 \right)^2
+ \left( \Phi_2^\dagger \Phi_2 \right)^2
+ \left( \Phi_3^\dagger \Phi_3 \right)^2 \right]
+ \lambda_2 \left( \Phi_4^\dagger \Phi_4 \right)^2
\no & &
+ \lambda_3 \left(
\Phi_1^\dagger \Phi_1\, \Phi_2^\dagger \Phi_2
+ \Phi_1^\dagger \Phi_1\, \Phi_3^\dagger \Phi_3
+ \Phi_2^\dagger \Phi_2\, \Phi_3^\dagger \Phi_3
\right)
\no & &
+ \lambda_4\, \Phi_4^\dagger \Phi_4
\left( \Phi_1^\dagger \Phi_1 + \Phi_2^\dagger \Phi_2
+ \Phi_3^\dagger \Phi_3 \right)
\no & &
+ \lambda_5 \left(
\left| \Phi_1^\dagger \Phi_2 \right|^2
+ \left| \Phi_1^\dagger \Phi_3 \right|^2
+ \left| \Phi_2^\dagger \Phi_3 \right|^2
\right)
\no & &
+ \lambda_6 \left( \left| \Phi_1^\dagger \Phi_4 \right|^2
+ \left| \Phi_2^\dagger \Phi_4 \right|^2
+ \left| \Phi_3^\dagger \Phi_4 \right|^2
\right)
\no & &
+ \left\{ \lambda_7 e^{i \zeta_7} \left[
\left( \Phi_1^\dagger \Phi_2 \right)^2
+ \left( \Phi_2^\dagger \Phi_3 \right)^2
+ \left( \Phi_3^\dagger \Phi_1 \right)^2
\right]
\right. \no & &
+ \lambda_8 e^{i \zeta_8} \left[
\left( \Phi_1^\dagger \Phi_4 \right)^2
+ \left( \Phi_2^\dagger \Phi_4 \right)^2
+ \left( \Phi_3^\dagger \Phi_4 \right)^2
\right]
\no & &
+ \lambda_9 e^{i \zeta_9} \left(
  \Phi_2^\dagger \Phi_3\, \Phi_1^\dagger
+ \Phi_3^\dagger \Phi_1\, \Phi_2^\dagger
+ \Phi_1^\dagger \Phi_2\, \Phi_3^\dagger
\right) \Phi_4
\no & & \left.
+ \lambda_{10} e^{i \zeta_{10}} \left(
  \Phi_3^\dagger \Phi_2\, \Phi_1^\dagger
+ \Phi_1^\dagger \Phi_3\, \Phi_2^\dagger
+ \Phi_2^\dagger \Phi_1\, \Phi_3^\dagger
\right) \Phi_4
+ {\rm H.c.} \right\},
\label{VA4}
\ea
where $\lambda_{1 \mbox{--} 10}$ are real.

\subsection{Vacuum} \label{vacuum}

In its particle content and symmetries,
hence in its Lagrangian,
the present model is almost identical to the one of Hirsch
\textit{et al.}~\cite{valle}.\footnote{The latter model has one extra
right-handed neutrino, invariant under $A_4$.}
The two models differ,
though,
in the assumed form of the vacuum state.\footnote{In reference~\cite{toorop}
a detailed study of the possible vacuum states of a model
with three Higgs doublets in an $A_4$ triplet was performed.
However,
in our model there is one extra,
$A_4$-invariant Higgs doublet---$\Phi_4$.
That extra doublet
changes things,
as we shall soon see,
mainly because of the presence in the scalar potential
of the extra terms with coefficients $\lambda_9$ and $\lambda_{10}$.
As a consequence,
the vacuum that we shall employ in this paper
does not exist in the model studied in reference~\cite{toorop}.}

We write the VEVs as
\be
\left\langle 0 \left| \phi_j^0 \right| 0 \right\rangle \equiv v_j
= \sqrt{V_j}\, e^{i \vartheta_j},
\ee
where the $\sqrt{V_j}$ are real and positive by definition.
Without loss of generality we set $\vartheta_4 = 0$.
We furthermore define
\ba
\chi_1 &=& \vartheta_1 - \vartheta_2 - \vartheta_3, \\
\chi_2 &=& \vartheta_2 - \vartheta_3 - \vartheta_1, \\
\chi_3 &=& \vartheta_3 - \vartheta_1 - \vartheta_2.
\ea
Then,
the vacuum potential is given by
\ba
\mathcal{V} &=&
\mu \left( V_1 + V_2 + V_3 \right)
+ \lambda_1 \left( V_1^2 + V_2^2 + V_3^2 \right)
\no & &
+ \left( \lambda_3 + \lambda_5 \right)
\left( V_1 V_2 + V_1 V_3 + V_2 V_3 \right)
\no & &
+ 2 \lambda_7 \left[
V_1 V_2 \cos{\left( \zeta_7 + \chi_2 - \chi_1 \right)}
+ V_2 V_3 \cos{\left( \zeta_7 + \chi_3 - \chi_2 \right)}
\right. \no & & \left.
+ V_3 V_1 \cos{\left( \zeta_7 + \chi_1 - \chi_3 \right)}
\right]
\no & &
+ 2 \lambda_8 V_4 \left[
V_1 \cos{\left( \zeta_8 + \chi_2 + \chi_3 \right)}
+ V_2 \cos{\left( \zeta_8 + \chi_3 + \chi_1 \right)}
\right. \no & & \left.
+ V_3 \cos{\left( \zeta_8 + \chi_1 + \chi_2 \right)}
\right]
\no & &
+ 2 \sqrt{V_1 V_2 V_3 V_4} \left\{
\lambda_9 \left[
\cos{\left( \zeta_9 + \chi_1 \right)}
+ \cos{\left( \zeta_9 + \chi_2 \right)}
+ \cos{\left( \zeta_9 + \chi_3 \right)}
\right]
\right. \no & & \left.
+ \lambda_{10} \left[
\cos{\left( \zeta_{10} + \chi_1 \right)}
+ \cos{\left( \zeta_{10} + \chi_2 \right)}
+ \cos{\left( \zeta_{10} + \chi_3 \right)}
\right]
\right\},
\ea
where
\ba
\mathcal{V} &=& \left\langle 0 \left| V \right| 0 \right\rangle
- \mu_2 V_4 - \lambda_2 V_4^2,
\\
\mu &=&  \mu_1 + \left( \lambda_4 + \lambda_6 \right) V_4.
\ea

The equations for vacuum stationarity are
\ba
0 = \frac{\partial \mathcal{V}}{\partial V_1} &=&
\mu + 2 \lambda_1 V_1
+ \left( \lambda_3 + \lambda_5 \right) \left( V_2 + V_3 \right)
\no & &
+ 2 \lambda_7 \left[
V_2 \cos{\left( \zeta_7 + \chi_2 - \chi_1 \right)}
+ V_3 \cos{\left( \zeta_7 + \chi_1 - \chi_3 \right)}
\right]
\no & &
+ 2 \lambda_8 V_4
\cos{\left( \zeta_8 + \chi_2 + \chi_3 \right)}
\no & &
+ \sqrt{\frac{V_2 V_3 V_4}{V_1}} \left\{
\lambda_9 \left[
\cos{\left( \zeta_9 + \chi_1 \right)}
+ \cos{\left( \zeta_9 + \chi_2 \right)}
+ \cos{\left( \zeta_9 + \chi_3 \right)}
\right]
\right. \no & & \left.
+ \lambda_{10} \left[
\cos{\left( \zeta_{10} + \chi_1 \right)}
+ \cos{\left( \zeta_{10} + \chi_2 \right)}
+ \cos{\left( \zeta_{10} + \chi_3 \right)}
\right]
\right\},
\label{nvher}
\\
0 = \frac{\partial \mathcal{V}}{\partial V_2} &=&
\mu + 2 \lambda_1 V_2
+ \left( \lambda_3 + \lambda_5 \right) \left( V_1 + V_3 \right)
\no & &
+ 2 \lambda_7 \left[
V_3 \cos{\left( \zeta_7 + \chi_3 - \chi_2 \right)}
+ V_1 \cos{\left( \zeta_7 + \chi_2 - \chi_1 \right)}
\right]
\no & &
+ 2 \lambda_8 V_4
\cos{\left( \zeta_8 + \chi_3 + \chi_1 \right)}
\no & &
+ \sqrt{\frac{V_1 V_3 V_4}{V_2}} \left\{
\lambda_9 \left[
\cos{\left( \zeta_9 + \chi_1 \right)}
+ \cos{\left( \zeta_9 + \chi_2 \right)}
+ \cos{\left( \zeta_9 + \chi_3 \right)}
\right]
\right. \no & & \left.
+ \lambda_{10} \left[
\cos{\left( \zeta_{10} + \chi_1 \right)}
+ \cos{\left( \zeta_{10} + \chi_2 \right)}
+ \cos{\left( \zeta_{10} + \chi_3 \right)}
\right]
\right\},
\label{uvhfg} \\
0 = \frac{\partial \mathcal{V}}{\partial V_3} &=&
\mu + 2 \lambda_1 V_3
+ \left( \lambda_3 + \lambda_5 \right) \left( V_1 + V_2 \right)
\no & &
+ 2 \lambda_7 \left[
V_1 \cos{\left( \zeta_7 + \chi_1 - \chi_3 \right)}
+ V_2 \cos{\left( \zeta_7 + \chi_3 - \chi_2 \right)}
\right]
\no & &
+ 2 \lambda_8 V_4
\cos{\left( \zeta_8 + \chi_1 + \chi_2 \right)}
\no & &
+ \sqrt{\frac{V_1 V_2 V_4}{V_3}} \left\{
\lambda_9 \left[
\cos{\left( \zeta_9 + \chi_1 \right)}
+ \cos{\left( \zeta_9 + \chi_2 \right)}
+ \cos{\left( \zeta_9 + \chi_3 \right)}
\right]
\right. \no & & \left.
+ \lambda_{10} \left[
\cos{\left( \zeta_{10} + \chi_1 \right)}
+ \cos{\left( \zeta_{10} + \chi_2 \right)}
+ \cos{\left( \zeta_{10} + \chi_3 \right)}
\right]
\right\},
\label{jbuiy} \\
0 = \frac{\partial \mathcal{V}}{\partial \chi_1} &=&
2 \lambda_7 V_1 \left[
V_2 \sin{\left( \zeta_7 + \chi_2 - \chi_1 \right)}
- V_3 \sin{\left( \zeta_7 + \chi_1 - \chi_3 \right)}
\right]
\no & &
- 2 \lambda_8 V_4 \left[
V_2 \sin{\left( \zeta_8 + \chi_3 + \chi_1 \right)}
+ V_3 \sin{\left( \zeta_8 + \chi_1 + \chi_2 \right)}
\right]
\no & &
- 2 \sqrt{V_1 V_2 V_3 V_4} \left[
\lambda_9 \sin{\left( \zeta_9 + \chi_1 \right)}
+ \lambda_{10} \sin{\left( \zeta_{10} + \chi_1 \right)}
\right],
\\
0 = \frac{\partial \mathcal{V}}{\partial \chi_2} &=&
2 \lambda_7 V_2 \left[
V_3 \sin{\left( \zeta_7 + \chi_3 - \chi_2 \right)}
- V_1 \sin{\left( \zeta_7 + \chi_2 - \chi_1 \right)}
\right]
\no & &
- 2 \lambda_8 V_4 \left[
V_3 \sin{\left( \zeta_8 + \chi_1 + \chi_2 \right)}
+ V_1 \sin{\left( \zeta_8 + \chi_2 + \chi_3 \right)}
\right]
\no & &
- 2 \sqrt{V_1 V_2 V_3 V_4} \left[
\lambda_9 \sin{\left( \zeta_9 + \chi_2 \right)}
+ \lambda_{10} \sin{\left( \zeta_{10} + \chi_2 \right)}
\right],
\label{hfyri} \\
0 = \frac{\partial \mathcal{V}}{\partial \chi_3} &=&
2 \lambda_7 V_3 \left[
V_1 \sin{\left( \zeta_7 + \chi_1 - \chi_3 \right)}
- V_2 \sin{\left( \zeta_7 + \chi_3 - \chi_2 \right)}
\right]
\no & &
- 2 \lambda_8 V_4 \left[
V_1 \sin{\left( \zeta_8 + \chi_2 + \chi_3 \right)}
+ V_2 \sin{\left( \zeta_8 + \chi_3 + \chi_1 \right)}
\right]
\no & &
- 2 \sqrt{V_1 V_2 V_3 V_4} \left[
\lambda_9 \sin{\left( \zeta_9 + \chi_3 \right)}
+ \lambda_{10} \sin{\left( \zeta_{10} + \chi_3 \right)}
\right].
\label{mbkig}
\ea
(The stationarity equation of $\left\langle 0 \left| V \right| 0 \right\rangle$
relative to variations of $V_4$ will be written down later.)
It is clear that solutions to equations~(\ref{nvher})--(\ref{mbkig})
with $v_2 = v_3$,
\textit{i.e.}\ with $V_2 = V_3$ and $\chi_2 = \chi_3$,
exist if and only if $\zeta_7 = 0$ or $\pi$,\footnote{Here
we depart from reference~\cite{kuhbock}
(see also reference~\cite{toorop}),
in which the Higgs doublet $\Phi_4$ was not present
and a solution with $\chi_2 - \chi_1 = \chi_1 - \chi_3$
was uncovered when $\zeta_7 \neq 0, \pi$.}
\textit{cf.}\ equations~(\ref{uvhfg}) and (\ref{jbuiy}),
(\ref{hfyri}) and (\ref{mbkig}).
From now on
\emph{we assume that a symmetry CP is present at the Lagrangian level},
which enforces $\zeta_7 = \zeta_8 = \zeta_9 = \zeta_{10} = 0$
(remember that $\lambda_{7 \mbox{--} 10}$ are real
and may be either positive or negative).
We may then assume that the vacuum has $v_2 = v_3$,
with
\ba
0 &=&
\mu + 2 \lambda_1 V_1
+ 2 \left( \lambda_3 + \lambda_5 \right) V_2
+ 4 \lambda_7 V_2 \cos{\left( \chi_2 - \chi_1 \right)}
\no & &
+ 2 \lambda_8 V_4 \cos{\left( 2 \chi_2 \right)}
+ \lambda_s V_2 \sqrt{\frac{V_4}{V_1}}
\left( \cos{\chi_1} + 2 \cos{\chi_2} \right),
\label{yvcre} \\
0 &=&
\mu + 2 \lambda_1 V_2
+ \left( \lambda_3 + \lambda_5 \right) \left( V_1 + V_2 \right)
+ 2 \lambda_7 \left[
V_2 + V_1 \cos{\left( \chi_2 - \chi_1 \right)} \right]
\no & &
+ 2 \lambda_8 V_4
\cos{\left( \chi_1 + \chi_2 \right)}
+ \lambda_s \sqrt{V_1 V_4} \left( \cos{\chi_1} + 2 \cos{\chi_2} \right),
\label{chdgt} \\
0 &=&
2 \lambda_7 V_1 \sin{\left( \chi_2 - \chi_1 \right)}
- 2 \lambda_8 V_4 \sin{\left( \chi_1 + \chi_2 \right)}
- \lambda_s \sqrt{V_1 V_4} \sin{\chi_1},
\label{mviyr}
\\
0 &=&
\lambda_7 V_1 V_2 \sin{\left( \chi_1 - \chi_2 \right)}
- \lambda_8 V_4
\left[ V_2 \sin{\left( \chi_1 + \chi_2 \right)}
+ V_1 \sin{\left( 2 \chi_2\right)} \right]
\no & &
- \lambda_s V_2 \sqrt{V_1 V_4} \sin{\chi_2},
\label{ndjir}
\ea
where $\lambda_s = \lambda_9 + \lambda_{10}$.

In order to get a feeling for what is at stake,
let us consider CP-conserving solutions to
equations~(\ref{yvcre})--(\ref{ndjir})
with $\chi_1 = \chi_2 = 0$.
(In our actual fits we shall always use spontaneously CP-breaking solutions;
this paragraph should be understood merely as an illustration
of the consequences of $\lambda_s \neq 0$.)
Then,
equations~(\ref{mviyr}) and~(\ref{ndjir}) are automatically satisfied
while equations~(\ref{yvcre}) and~(\ref{chdgt}) read
\ba
2 \lambda_1 V_1
+ 2 \lambda_m V_2
+ 3 \lambda_s V_2 \sqrt{\frac{V_4}{V_1}}
&=& - \bar \mu,
\label{mvkit} \\
2 \lambda_1 V_2
+ \lambda_m \left( V_1 + V_2 \right)
+ 3 \lambda_s \sqrt{V_1 V_4}
&=& - \bar \mu,
\label{uiyre}
\ea
where
\ba
\lambda_m &=& \lambda_3 + \lambda_5 + 2 \lambda_7,
\\
\bar \mu &=& \mu + 2 \lambda_8 V_4.
\ea
There is a solution to equations~(\ref{mvkit})
and~(\ref{uiyre}) with $V_1 = V_2$.
If and only if $\lambda_s \neq 0$,\footnote{One also needs to assume
that some combinations of the coefficients
have the appropriate signs.}
there is also a solution with (in general) $V_1 \neq V_2$,
\ba
\sqrt{V_4} &=& \frac{2 \lambda_1 - \lambda_m}{3 \lambda_s} \sqrt{V_1},
\\
V_2 &=& \frac{- \bar \mu - 2 \lambda_1 V_1}{2 \lambda_1 + \lambda_m}.
\ea
Note that $\lambda_s \neq 0$ is crucial for the existence
of this solution.
The presence in the potential
of the terms with coefficients $\lambda_9$ and $\lambda_{10}$
leads to the existence of stationarity points with $V_2 = V_3 \neq V_1$.

\subsection{The neutrino mass matrix}

We assume that the stationarity point of the scalar potential
found in the previous subsection,
with $v_2 = v_3$,
is indeed the \emph{global minimum} of the potential,\footnote{In our
numerical work we have demonstrated that the stationarity points
that we employ are local minima of the potential,
\textit{i.e.}\
we have checked that
all the corresponding scalar squared masses are positive.
The demonstration that those local minima
are \emph{the global minimum}
of the potential would be much more involved
and is outside the scope of the present paper.}
\textit{i.e.}\ we assume that \emph{the vacuum state has}
\be
v_2 = v_3.
\ee
Then,
according to equations~(\ref{M1}) and~(\ref{uvhqw}),
\be
M = - \frac{1}{m} \left( \begin{array}{ccc}
a^2 r & a b s & a c s \\
a b s & b^2 s & b c r \\
a c s & b c r & c^2 s
\end{array} \right).
\label{M2}
\ee
where
\ba
r &=& {v_1^\ast}^2 + 2 {v_2^\ast}^2,
\\
s &=& {v_1^\ast}^2 - {v_2^\ast}^2
\ea
are in general complex and have unrelated phases.
The Yukawa couplings $a$, $b$, and $c$ must be taken 
real in equation~(\ref{M2}),
since we have assumed CP symmetry at the Lagrangian level.

The light-neutrino mass matrix in equation~(\ref{M2}) obeys
the two rephasing-invariant constraints
in equations~(\ref{constraint1}) and~(\ref{constraint2}).
Another remarkable feature of the matrix in equation~(\ref{M2})
is that it preserves its form when it is inverted,
\textit{i.e.}\ $M^{-1}$ is of the same form as $M$
and also satisfies the constraints~(\ref{constraint1}) and~(\ref{constraint2}).

\section{Fits} \label{predictions}

In the basis where the charged-lepton mass matrix is diagonal,
the neutrino Majorana mass matrix $M$
is bi-diagonalized by the lepton mixing matrix $U$:
\be
M = U^\ast D U^\dagger,
\label{M}
\ee
where
\be
D = \mathrm{diag}
\left( m_1,\, m_2 e^{- i \chi_{21}},\, m_3 e^{- i \chi_{31}} \right).
\ee

In our numerical work we have
inputted
random initial values
of the neutrino masses,
of the phases $\chi_{21}$ and $\chi_{31}$,
and of the parameters of $U$,
within their respective $3 \sigma$ intervals
given in table~\ref{data} (parameters which are not in that table
were taken free).\footnote{We have also performed a fit
of our model to the phenomenological data
of reference~\cite{fogli}---with the values given there
for the `new reactor data'---and we have found that
our model is compatible with those data
with about the same level of stress as the one registered
in the fit presented here.}
We computed the matrix $M$ by using equation~(\ref{M}).
We have then step-by-step adjusted the input parameters
in order to eventually fit
the constraints~(\ref{constraint1}) and~(\ref{constraint2}).

For each of the fits thus obtained,
we have calculated
\be
\frac{r}{s} = \frac{M_{ee} M_{\mu\mu}}{\left( M_{e\mu} \right)^2}
\ee
and therefrom obtained
\be
\frac{v_2}{v_1} = \sqrt{\frac{r^\ast - s^\ast}{r^\ast + 2 s^\ast}}
= \sqrt{\frac{V_2}{V_1}}\, \exp{\left( i\ \frac{\chi_2 - \chi_1}{2} \right)}.
\ee
We have further inputted random values of $V_1$, $V_4$, $\chi_1$,
and of the parameters $\lambda_{1,2,3,4,6,7,9}$.
By using the four stationarity equations~(\ref{yvcre})--(\ref{ndjir})
together with
\ba
0 = \frac{\partial \left\langle 0 \left| V \right| 0 \right\rangle}
{\partial V_4} &=&
\mu_2 + 2 \lambda_2 V_4
+ \left( \lambda_4 + \lambda_6 \right) \left( V_1 + 2 V_2 \right)
\no & &
+ 2 \lambda_8 \left[
V_1 \cos{\left( 2 \chi_2 \right)}
+ 2 V_2 \cos{\left( \chi_1 + \chi_2 \right)}
\right]
\no & &
+ \lambda_s \sqrt{\frac{V_1}{V_4}}\, V_2
\left( \cos{\chi_1} + 2 \cos{\chi_2} \right),
\ea
we have calculated the remaining five parameters of the potential,
\textit{i.e.}\ $\lambda_{5,8,10}$ and $\mu_{1,2}$.
We have then computed the neutral and charged-scalar mass matrices
at this stationarity point and discarded the point
whenever any of the physical scalars displayed
a negative squared mass,
\textit{i.e.}\ we have made sure that the stationarity point
is indeed a local minimum of the potential.
We have also checked that the correct number of Goldstone bosons
is present at each local minimum.

We have discovered that
many of the local minima of the potential thus found
display low-mass scalars;
indeed,
\emph{all} the local minima feature
\emph{at least one low-mass physical neutral scalar}.
In order to contain this problem of low-mass scalars,
we have discarded any minima in which either a charged scalar
or more than one neutral scalar has mass smaller than $100~\mathrm{GeV}$.
The masses were normalized through
\be
V_1 + 2 V_2 + V_4 = \left( 174~\mathrm{GeV} \right)^2.
\label{hcuer}
\ee

\begin{figure}
\begin{center}
\mbox{\epsfig{file=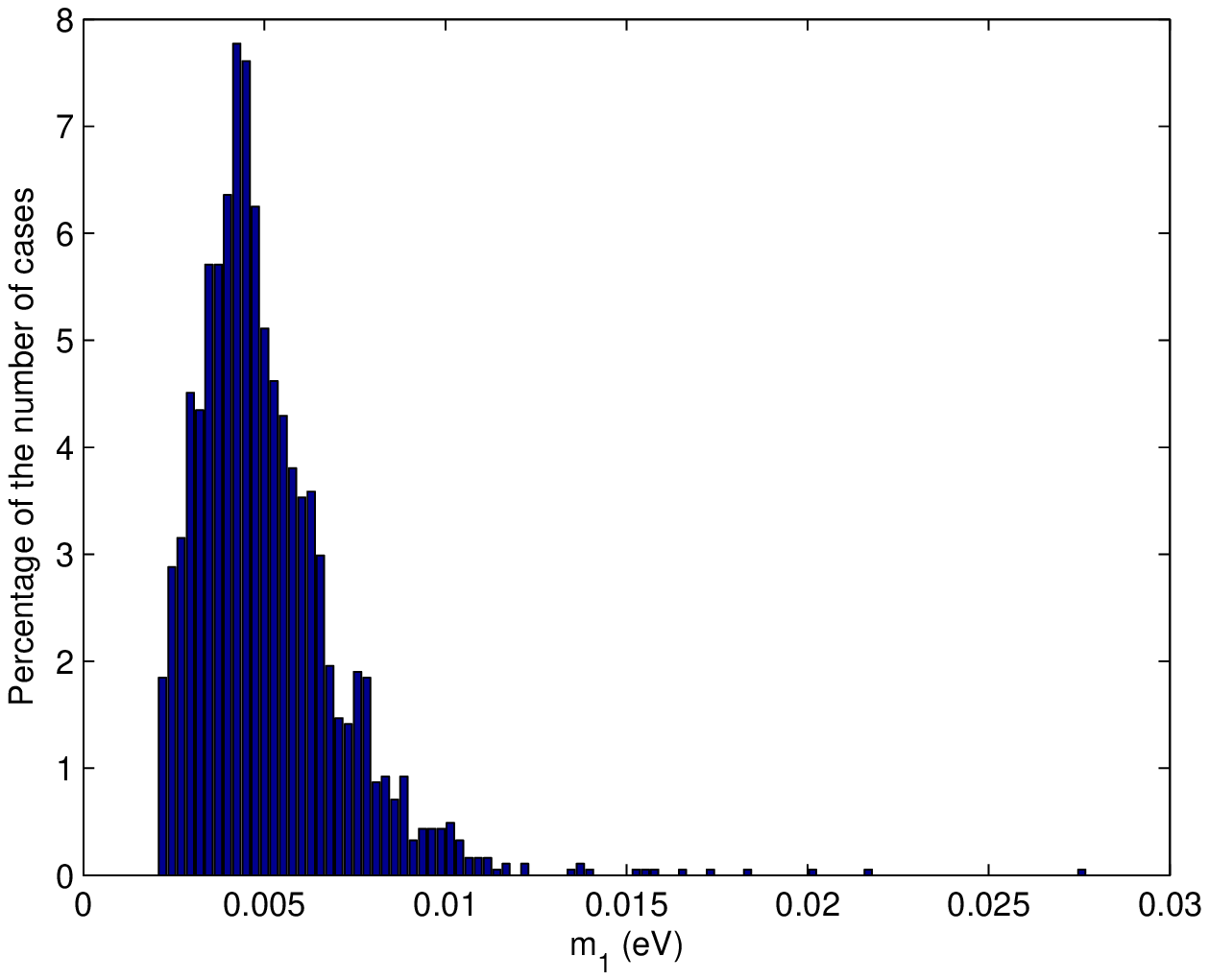,
width=0.8\textwidth,
height=10cm}}
\end{center}
\caption{Histogram displaying the distribution of the lowest neutrino mass,
$m_1$,
among the points with a normal neutrino mass spectrum.
In our set of such points,
$0.0021~\mathrm{eV} \le m_1 \le 0.0277~\mathrm{eV}$.}
\label{normalmass}
\end{figure}

\begin{figure}
\begin{center}
\mbox{\epsfig{file=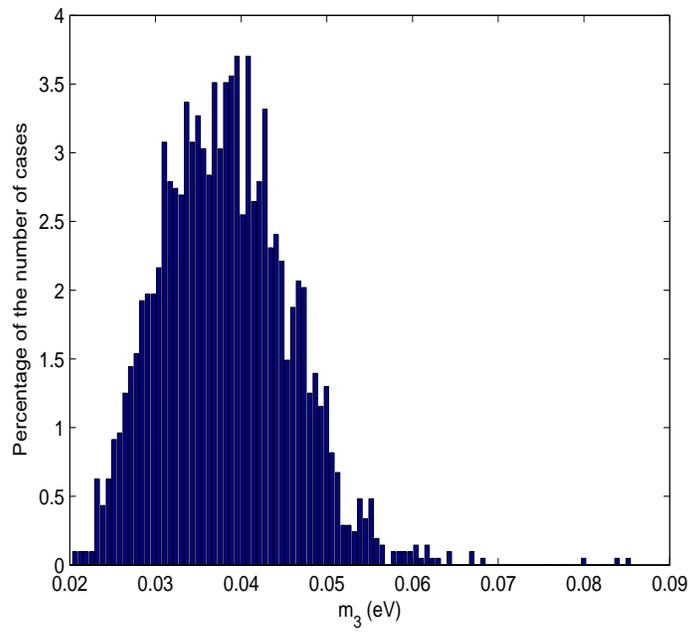,
width=0.75\textwidth,
height=9cm}}
\end{center}
\caption{Histogram of the distribution of the lowest neutrino mass,
$m_3$,
in points with an inverted neutrino mass spectrum.
For all those points
$0.0203~\mathrm{eV} \le m_1 \le 0.0855~\mathrm{eV}$.}
\label{invertedmass}
\end{figure}

In this way we have constructed two sets of points,
of approximately 1,800 points each,
one of them with normal and the other one with inverted neutrino mass spectra,
obeying the constraints~(\ref{constraint1}) and~(\ref{constraint2})
and which are local minima of the scalar potential
with all the physical scalars but one
having a high mass.
These are the points that we next display in various scatter plots.
In all figures but for figures~\ref{normalmass} and~\ref{invertedmass},
blue (red) points are those with a normal (inverted) neutrino mass spectrum.

We first focus on the predictions of our model
for the absolute neutrino mass scale.
In figures~\ref{normalmass} and~\ref{invertedmass}
we display histograms of the lowest neutrino mass
for our sets of points.
One sees that the neutrino masses tend to be lower
when the mass spectrum is normal;
points with an inverted spectrum sometimes display neutrino masses
which are almost degenerate,
with $m_3 \sim \sqrt{m_1^2 - m_3^2} \approx 0.05~\mathrm{eV}$.
In contrast,
points with a normal neutrino mass spectrum usually display
a markedly hierarchical spectrum,
\textit{i.e.}\ one with $m_1 \ll \sqrt{m_3^2 - m_1^2}$.

\begin{figure}
\begin{center}
\mbox{\epsfig{file=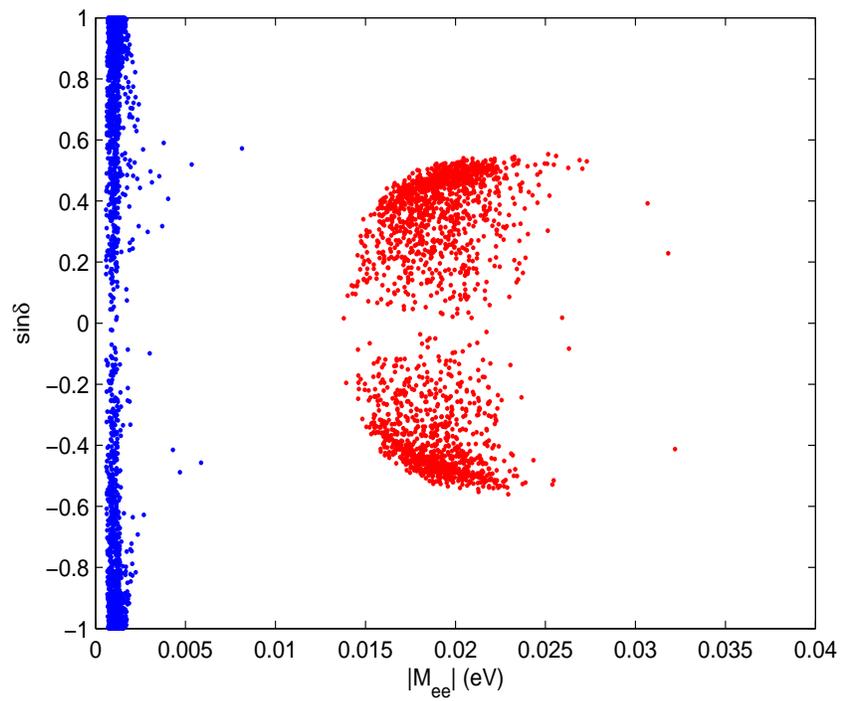,
width=0.9\textwidth,
height=10cm}}
\end{center}
\caption{Scatter plot of the Dirac phase against $m_{2\beta 0\nu}$.
Here and in the following figures,
points with a normal neutrino mass spectrum are marked blue,
those with an inverted spectrum are marked red.}
\label{deltabetabeta}
\end{figure}

In figure~\ref{deltabetabeta}
the phase $\delta$ is plotted against the mass term $\left| M_{ee} \right|$;
the latter is the quantity relevant for neutrinoless double-$\beta$ decay.
One sees once again that points with an inverted spectrum
display higher masses---$\left| M_{ee} \right|$ is there
typically $0.02~\mathrm{eV}$ but it is much lower
for points with normal neutrino spectra.
In our model there is no prediction for $\delta$
in the case of a normal spectrum,
while $\left| \sin{\delta} \right| \lesssim 0.5$
when the neutrino mass spectrum is inverted.

\begin{figure}
\begin{center}
\mbox{\epsfig{file=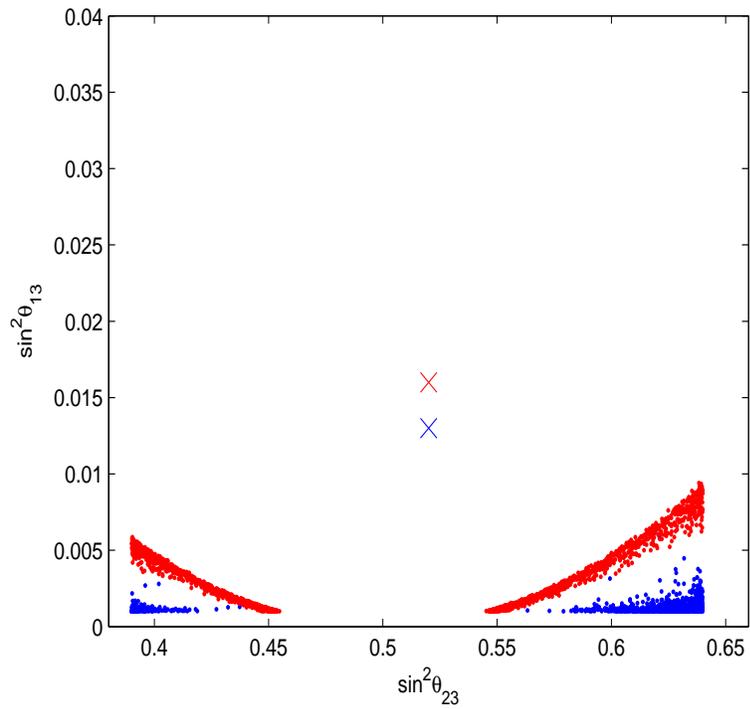,
width=0.8\textwidth,
height=10cm}}
\end{center}
\caption{Scatter plot of the reactor angle $\theta_{13}$
against the atmospheric angle $\theta_{23}$.
The crosses mark the best-fit points in table~\ref{data}.}
\label{tetateta}
\end{figure}

Figure~\ref{tetateta} gives the prediction of our model
for the reactor mixing angle $\theta_{13}$.
Notice that,
in our search for fits,
we have enforced the condition that all the observables
be within their respective $3 \sigma$ intervals
displayed in the third column of table~\ref{data};
this explains the blank areas in the lower part
and in both sides of figure~\ref{tetateta}.
One sees in that figure
that our model predicts
a very small $\theta_{13}$---$\sin^2{\theta_{13}}$ is always smaller
than one half the current best-fit values.
Since the experimental indications for a non-zero $\theta_{13}$
are still at an early stage,
and the precise value of that parameter is still debatable,
we do not consider this tension with the current bets-fit values
to be too bad
for our model.\footnote{Very recently,
the first data from the Daya Bay experiment have appeared~\cite{dayabay}
and they confirm a rather high $\theta_{13}$,
thus worsening the status of our fit.}
The fact that figure~\ref{tetateta} displays the atmospheric mixing angle
as being far from its `maximal' value $45^\circ$
is just a consequence of the fact that we enforce $s_{13}^2 \ge 0.001$
on our fits; indeed, in our model $\theta_{23} \to 45^\circ$
as $\theta_{13} \to 0$---our model is compatible
with $\mu$--$\tau$ interchange symmetry in the neutrino mass matrix---and
a phenomenological lower limit on $\theta_{13}$
implies in our model a lower limit on $\left| \theta_{23} - 45^\circ \right|$.

\begin{figure}
\begin{center}
\mbox{\epsfig{file=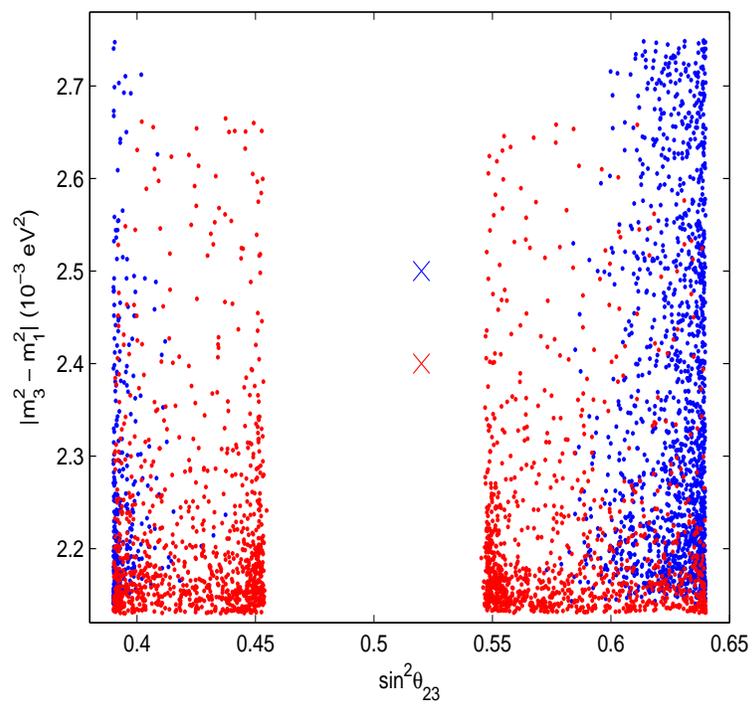,
width=0.8\textwidth,
height=10cm}}
\end{center}
\caption{Fit of our model to the atmospheric-neutrino oscillations.}
\label{atm}
\end{figure}

Figure~\ref{atm} is a scatter plot
of $\left| m_3^2 - m_1^2 \right|$ against $\theta_{23}$ in our model.
One sees that our model
tolerates well any phenomenological value of $\left| m_3^2 - m_1^2 \right|$,
but that cases with a normal neutrino mass spectrum
tend to have a worse fit of $\theta_{23}$
than those with an inverted spectrum.

\begin{figure}
\begin{center}
\mbox{\epsfig{file=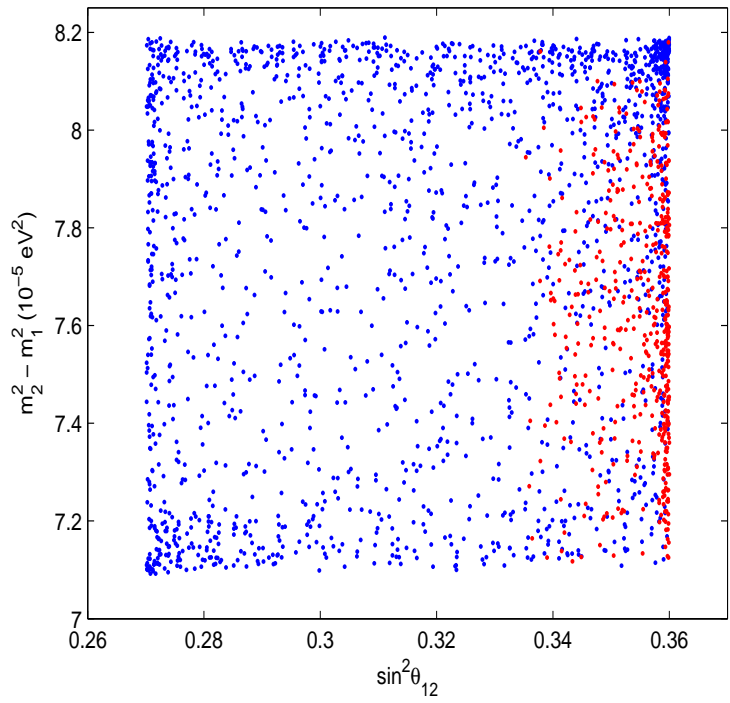,
width=0.8\textwidth,
height=10cm}}
\end{center}
\caption{Fit of our model to the solar-neutrino oscillations.}
\label{sol}
\end{figure}

Figure~\ref{sol} is the scatter plot
of $m_2^2 - m_1^2$ against the solar mixing angle.
In this case solutions with a normal neutrino mass spectrum
are sometimes excellent at fitting the phenomenological $\theta_{12}$;
those with an inverted spectrum display some tension with the data,
since they usually have $\theta_{12}$ quite larger than its best-fit value.

\begin{figure}
\begin{center}
\mbox{\epsfig{file=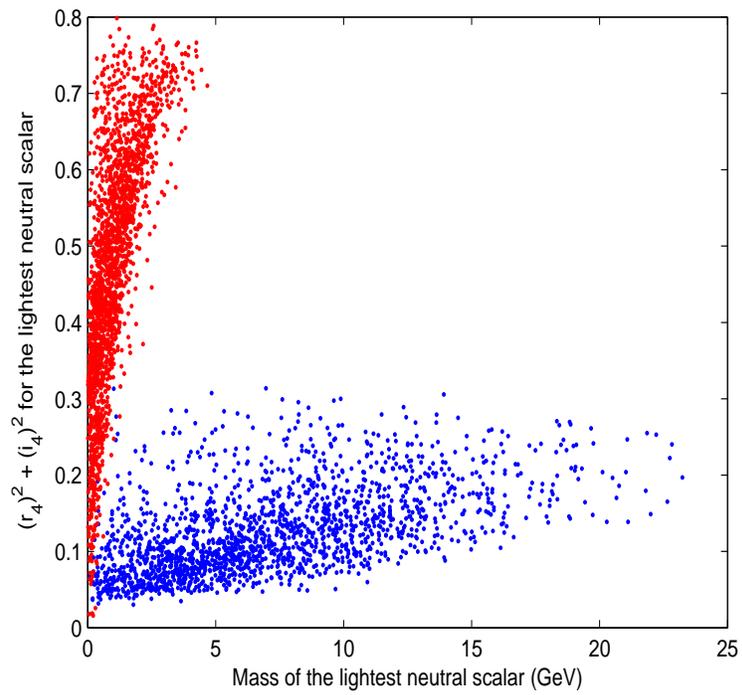,
width=0.8\textwidth,
height=10cm}}
\end{center}
\caption{The $\Phi_4$ component of the lightest neutral scalar
plotted against the mass of that scalar.}
\label{scalar}
\end{figure}

We next turn to the low-mass neutral scalar
which is an---indirect---prediction of our model.
We remind the reader
that we have enforced on our points the condition that
all scalars but one neutral one have mass larger than $100~\mathrm{GeV}$.
Let $m_\mathrm{light}$ denote the mass of the lightest physical neutral scalar.
In figure~\ref{scalar} we have plotted
$m_\mathrm{light}$.
We see in that figure that
$m_\mathrm{light} \lesssim 25~\mathrm{GeV}$
for a normal neutrino mass spectrum,
but
$m_\mathrm{light}$
is much lower---$5~\mathrm{GeV}$ or less---in the case
of an inverted neutrino mass spectrum.\footnote{If one wants to avoid a
very-low-mass scalar,
then one may add to the potential quadratic terms
which break the $A_4$ symmetry softly~\cite{toorop2}.
It is possible to find a soft breaking that preserves $v_2 = v_3$.}

Let $S_\mathrm{light}$ denote the light-scalar field;
we write it as
\be
S_\mathrm{light} = \sum_{j=1}^4 \left[
r_j\, \mathrm{Re} \left( \phi_j^0 - v_j \right)
+ i_j\, \mathrm{Im} \left( \phi_j^0 - v_j \right) \right],
\ee
where the $r_j$ and the $i_j$ are real and are normalized through
\be
\sum_{j=1}^4
\left[ \left( r_j \right)^2 + \left( i_j \right)^2 \right] = 1.
\ee
In the vertical axis of figure~\ref{scalar}
we display the coupling of the low-mass scalar to $\Phi_4$.
This is especially relevant since $\phi_{1,2,3}^0$ couple to neutrinos;
only $\phi_4^0$ couples to $\bar \alpha_L \alpha_R$.
Thus, if both $r_4$ and $i_4$ are small,
then $S_\mathrm{light}$ has suppressed couplings
to the charged leptons and may become invisible.
One notices in figure~\ref{scalar}
that cases with a normal neutrino mass spectrum
usually display both a larger $m_\mathrm{light}$
and smaller $r_4$ and $i_4$ than cases with an inverted spectrum.

\begin{figure}
\begin{center}
\mbox{\epsfig{file=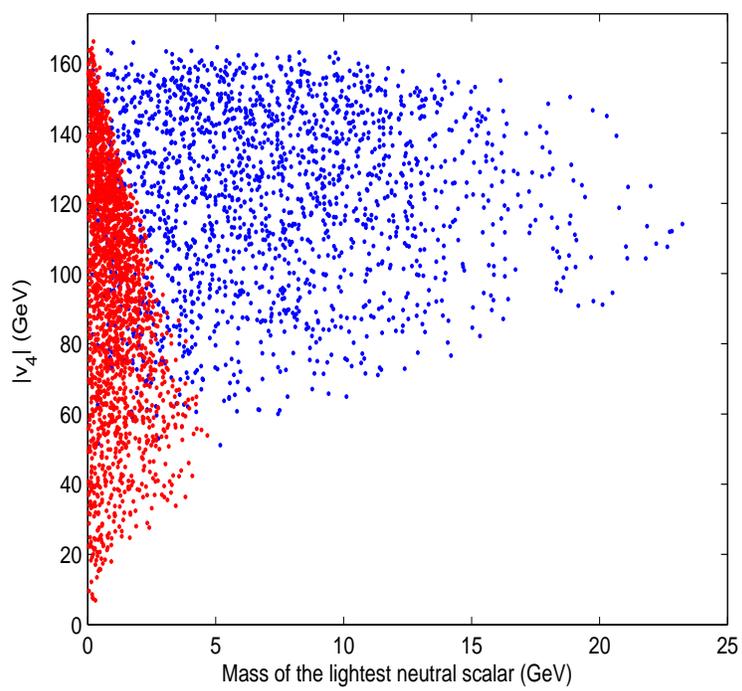,
width=0.8\textwidth,
height=10cm}}
\end{center}
\caption{The vacuum expectation value of $\phi_4^0$
plotted against the mass of the lightest neutral scalar.}
\label{scalar2}
\end{figure}

Moreover,
in our model,
just as in the Standard Model,
the coupling of the neutral scalars to the charged leptons
is suppressed by the charged-lepton mass,
\textit{cf.}\ equation~(\ref{jbvut}).
Now,
as seen in figure~\ref{scalar2},
almost all our points have
$\left| v_4 \right| \gtrsim\, 20~\mathrm{GeV}$.\footnote{As a matter of fact,
there are points for which $\left| v_4 \right| = \sqrt{V_4}$
almost saturates equation~(\ref{hcuer}).}
Therefore,
the coupling of $\phi_4^0$ to either $\bar e_L e_R$ or $\bar \mu_L \mu_R$
is always extremely small.

A possible discovery channel of the light scalar
would have been through the process
$e^+ e^- \to Z^0 S_\mathrm{light}$ at LEP.
The LEP bound on this process only extends down
to $m_\mathrm{light} \gtrsim 12~\mathrm{GeV}$,
\textit{cf.}\ table~14 of reference~\cite{lep}.
Another possible discovery channel for the light scalar
would have been $e^+ e^- \to Z^\ast \to S_\mathrm{light} S^\prime$,
where $S^\prime$ is any neutral scalar
heavier than $S_\mathrm{light}$---there are six possible $S^\prime$ in our model.
In order to investigate these possibilities we have computed,
for each of our points and for all seven physical scalars $S_i$,
the strengths of the couplings $Z Z S_i$
and $Z S_i S_j$.
We have compared those strengths with the data
in tables~14 and~19 of reference~\cite{lep},
which are the LEP bounds on $e^+ e^- \to Z S_i$
and on $e^+ e^- \to S_i S_j$,
respectively, assuming that both $S_i$ and $S_j$
decay exclusively into $\tau^+ \tau^-$; the second bound is effective for $m_i
+ m_j \lesssim 200$ GeV,
the LEP kinematic limit.
We have found that only a small percentage of our points
(about 16\% in the normal case,
10\% in the inverted case)
can be eliminated in this way.
This is because,
for most of our points,
either the $S_i$ are much too heavy,
thus kinematically evading the LEP bound,
or the $S_i$ have much too small couplings
(in many cases,
zero for all practical purposes)
to $Z S_j$;
moreover, even when they are with the kinematic
limits of LEP, the 
$S_i$ almost always have a minuscule coupling to $Z Z$.
Remaking our plots by using only the points that have survived these tests,
we have found that they look undistinguishable from the ones presented
in figures 1--6.
We remark that our tests are in all likelihood much too strict,
since usually in our model the neutral scalars
will not decay exclusively into $\tau^+ \tau^-$.

Our $S_\mathrm{light}$ is not necessarily produced at the LHC
through gauge-boson fusion,
since it does not need to couple to the top quark---we remark that
in this paper we have not specified the quark Yukawa couplings,
which might even necessitate the addition to the model
of extra Higgs doublets.

\section{Summary} \label{conclusions}

In this paper we have discovered that
the $A_4$-symmetric renormalizable scalar potential
for an $SU(2) \times U(1)$ gauge theory
with one $A_4$ triplet of Higgs doublets,
together with one $A_4$-invariant doublet,
allows (local) minima for which two of the Higgs doublets
in the $A_4$ triplet have equal VEVs.
We have made use of such minima in a specific seesaw model
with $A_4$ horizontal symmetry in the lepton sector.
We have thus obtained a renormalizable model
which makes the predictions~(\ref{constraint1}) and~(\ref{constraint2})
for the (effective) light-neutrino Majorana mass matrix $M$
in the basis where the charged-lepton mass matrix is diagonal.
We have shown that those predictions are compatible with
the phenomenological data on neutrino masses and mixings,
irrespective of whether the neutrino mass spectrum is normal or inverted.
Remarkably,
we have found that in all such cases the scalar potential
turns out to lead to (at least) one very light neutral scalar,
with mass not larger than 25 (5)~GeV
in the cases with normal (inverted) neutrino mass spectrum.


\paragraph{Acknowledgements:}
The work of L.L.\ is funded by the Portuguese
{\it Fun\-da\-\c c\~ao para a Ci\^encia e a Tecnologia} (FCT)
through FCT unit 777 and through the projects CERN/FP/116328/2010,
PTDC/FIS/098188/2008,
and PTDC/FIS/117951/2010,
and also by the Marie Curie Initial Training Network ``UNILHC''
PITN-GA-2009-237920.
The work of P.F.\ is supported in part by the Portuguese
\textit{Fun\-da\-\c{c}\~{a}o para a Ci\^{e}ncia e a Tecnologia} (FCT)
under contract PTDC/FIS/117951/2010,
by the FP7 Reintegration Grant n.~PERG08-GA-2010-277025,
and by PEst-OE/FIS/UI0618/2011.


\newpage


\end{document}